\documentclass[fleqn,10pt]{wlscirep}
\title{Quantum Phase Transition and Entanglement in Topological Quantum Wires}

\author[1,2,*]{Jaeyoon Cho}
\author[3]{Kun Woo Kim}
\affil[1]{Asia Pacific Center for Theoretical Physics, Pohang 37673, Korea}
\affil[2]{Department of Physics, POSTECH, Pohang 37673, Korea}
\affil[3]{School of Physics, Korea Institute for Advanced Study, Seoul 02455, Korea}

\affil[*]{choooir@gmail.com}

\begin{abstract}
We investigate the quantum phase transition of the Su-Schrieffer-Heeger (SSH) model by inspecting the two-site entanglements in the ground state. It is shown that the topological phase transition of the SSH model is signified by a nonanalyticity of local entanglement, which becomes discontinuous for finite even system sizes, and that this nonanalyticity has a topological origin. Such a peculiar singularity has a universal nature in one-dimensional topological phase transitions of noninteracting fermions. We make this clearer by pointing out that an analogous quantity in the Kitaev chain exhibiting the identical nonanalyticity is the local electron density. As a byproduct, we show that there exists a different type of phase transition, whereby the pattern of the two-site entanglements undergoes a sudden change. This transition is characterised solely by quantum information theory and does not accompany the closure of the spectral gap. We analyse the scaling behaviours of the entanglement in the vicinities of the transition points.
\end{abstract}

\newcommand{\ket}[1]{|#1\rangle}
\newcommand{\avr}[1]{\langle#1\rangle}

\begin{document}

\flushbottom
\maketitle

\thispagestyle{empty}

\section*{Introduction}

Quantum phase transition is one of the pillars underpinning condensed matter physics~\cite{sac11}. Conventional wisdom states that different quantum phases are generally discriminated in terms of the symmetry carried by the ground state or other features that have an underlying topological interpretation~\cite{wen04}. The former is described by local order parameters associated with the symmetries and the latter by topological orders, which are nowadays classified into intrinsic and symmetry-protected ones~\cite{wen90,kit01,wen04,sch08,gu09,che10,fid10,has10}. In both cases, a continuous transition between different phases is mediated by a spectrally gapless critical point, in the vicinity of which thermodynamic quantities exhibit scaling behaviours classified into universality classes~\cite{fis98}.

Modern understanding of quantum phase transition has been significantly enriched by incorporating the concept of entanglement~\cite{ost02,osb02,vid03,kit06,lev06,li08,che10,fid10,ali12}. Quantum phases are determined after all by the way how different particles or different parties in the system are mutually related. In this sense, it is natural to expect that entanglement would bear the fingerprint of the quantum phase. This perspective is especially powerful in the study of topological orders, which are a purely quantum effect. For example, states with an intrinsic topological order have a long-range entanglement and a nonzero topological entanglement entropy~\cite{kit06,lev06,che10}. Symmetry-protected topological orders are signified by a degenerate entanglement spectrum~\cite{fid10}. Topological quantum phase transition would then be thought of as a rearrangement of the pattern of entanglement.

While it is a common practice to study macroscopic bipartite entanglements in topological phases, they apparently reveal only a partial aspect of many-body entanglement and, on the practical side, are hardly accessible in experiments. It is thus worthwhile to carry out a more detailed inspection of the many-body entanglement for a deeper understanding of topological phases. In particular, when it comes to the aspect of phase transition, local entanglement may be enough to gain information on the critical singularities, as is suggested by earlier works on symmetry-breaking quantum phase transitions in Heisenberg spin chains~\cite{ost02,osb02}. If then, an interesting question is how its singular nature differs from that of the symmetry-breaking transitions. Besides, from the viewpoint that different quantum phases are imprinted in different patterns of entanglement, to examine many-body entanglement in topological models is an interesting problem in its own right.

\begin{figure}
\center\includegraphics[width=0.6\columnwidth]{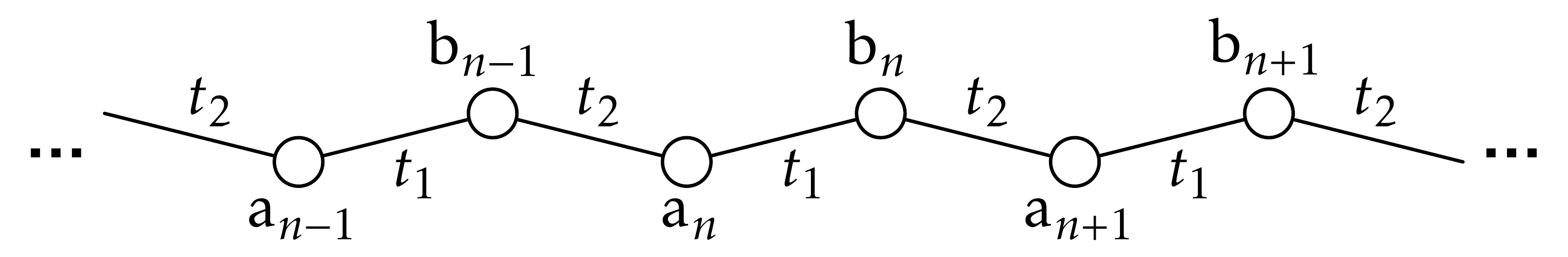}
\caption{Su-Schrieffer-Heeger model. A unit cell consists of two sites. The hopping rates are set to be $t_{1}=1-\lambda$ and $t_{2}=1+\lambda$ for $0\le\lambda\le1$.}
\label{fig:model}
\end{figure}

The aim of this paper is to investigate the quantum phase transition of one-dimensional topological models in terms of the two-site entanglements, namely, the concurrences, in the ground state~\cite{hor09}. As a prototypical model, we consider the Su-Schrieffer-Heeger (SSH) model on a one-dimensional lattice, as shown in Figure~\ref{fig:model}~\cite{su79}. The system has $N$ unit cells, each consisting of two sites $\{\text{a}_{n},\text{b}_{n}\}$. The model Hamiltonian is given by
\begin{equation}
H=\sum_{n=1}^{N}\left(t_{1}a_{n}^{\dagger}b_{n}+t_{2}b_{n}^{\dagger}a_{n+1}+\text{H.c.}\right),
\label{eq:model}
\end{equation}
where $\{a_{n},b_{n}\}$ denote the fermion operators for the $n$-th unit cell and the periodic boundary condition $\{a_{N+1},b_{N+1}\}=\{a_{1},b_{1}\}$ is taken. We take the hopping rates 
\begin{equation}
t_{1}=1-\lambda,\quad t_{2}=1+\lambda
\end{equation}
to have a single control parameter $\lambda\in[-1,1]$. This model has been well studied in the context of the band topology~\cite{sch08,has10}. It is in a topological phase for $\lambda>0$ and in a trivial phase for $\lambda<0$ (this distinction of the phase, of course, depends on the choice of the unit cell). 

We first derive an analytic formula for the concurrence between any two sites and show that in the thermodynamic limit, the first derivative of the concurrence between adjacent sites with respect to $\lambda$ diverges logarithmically at the critical point $\lambda_{0}=0$, the exact form of which is also derived. This result is similar to the case of the symmetry-breaking quantum phase transition in the Heisenberg spin chain~\cite{ost02,osb02}. However, due to the topological origin, there exists an interesting difference: for finite even $N$, the concurrence is discontinuous at $\lambda=\lambda_{0}$ with a gap inversely proportional to $N$, while it remains analytic for odd $N$. This feature contrasts with the case of symmetry-breaking quantum phase transitions wherein the nonanalyticity appears only in the thermodynamic limit. We provide a geometric interpretation of that, directly relating the singularity with the change of the band topology. This phenomenon has a universal nature in one-dimensional topological phase transitions of noninteracting fermions. As an example, we show that for the Kitaev chain~\cite{kit01}, the local electron density is an analogous quantity exhibiting the identical nonanalyticity at the critical point: its first derivative, i.e., the local compressibility, diverges logarithmically in the thermodynamic limit and it is discontinuous for finite even system sizes.

Another finding we present in this paper is that there exists a different type of phase transition in the SSH model whereby the many-body entanglement of the ground state undergoes a sudden change in the following sense. As a means to characterise the many-body entanglement, we represent the pairwise pattern of all the concurrences as a (simple) graph, where each edge means the existence of entanglement, i.e., a nonzero concurrence, between the two vertices (sites). We call this graph an ``entangled graph'', following Ref.~\cite{ple03}. If two many-body states have different entangled graphs, we will regard them as having different classes of many-body entanglement and hence belonging to different phases. This kind of characterisation of many-body entanglement has a relatively long tradition~\cite{dur00,bri01,ple03}. In the present work, our particular motivation is coming from the fact that at two extreme phases $\lambda=\pm1$, the ground state is dimerised in such a way that either two sites in every unit cell form a singlet ($\lambda=-1$) or every adjacent pair of sites across unit cells form a singlet ($\lambda=+1$), the entanglement of which can be naturally represented by the entangled graphs shown in Figure~\ref{fig:phase}. The problem now is then to figure out how the entangled graph for $\lambda\in(-1,1)$ interpolates between the two. It turns out that at $\lambda_{\pm}\simeq\pm0.138$, the entangled graph changes suddenly along with a nonanalyticity of the concurrence. This singularity is reminiscent of the phenomenon called the entanglement sudden death~\cite{yu09}. We emphasise, however, that this transition should not be confused with conventional quantum phase transitions because it has nothing to do with a nonanalyticity of the ground-state wavefunction itself: the nonanalytic behaviour comes from the way the entanglement is defined and quantified. A further remark on this transition will be made later. In what follows, ``quantum phase transition'' will refer only to the transition at $\lambda=\lambda_{0}$. 

\section*{Results}

\subsection*{Reduced density matrices in the Su-Schrieffer-Heeger model} 

It is convenient to switch to the momentum space, in which the Hamiltonian~\eqref{eq:model} takes the Bogoliubov-de Gennes form:
\begin{equation}
H=\sum_{k\in\text{BZ}}\phi_{k}^{\dagger}\left[\vec{h}(k)\cdot\vec{\sigma}\right]\phi_{k},
\label{eq:hamiltonian}
\end{equation}
where $\phi_{k}=(a_{k}~b_{k})^{T}$, $\vec{\sigma}=\sigma_{x}\hat{x}+\sigma_{y}\hat{y}+\sigma_{z}\hat{z}$ is the vector composed of Pauli matrices, and $\vec{h}(k)$ is given by
\begin{equation}
\vec{h}(k)=(t_{1}+t_{2}\cos k)\hat{x} +(t_{2}\sin k)\hat{y}.
\label{eq:vector}
\end{equation}
Our aim is to obtain the concurrences between all pairs of lattice sites, which we denote by $\mathcal{C}(\text{a}_{n},\text{b}_{m})$, $\mathcal{C}(\text{a}_{n},\text{a}_{m})$, and $\mathcal{C}(\text{b}_{n},\text{b}_{m})$. For this, we need to obtain the corresponding reduced density matrices $\rho_{\text{a}_{n}\text{b}_{m}}$, $\rho_{\text{a}_{n}\text{a}_{m}}$, and $\rho_{\text{b}_{n}\text{b}_{m}}$. To this end, we follow the method presented in Ref.~\cite{fid10}, which we also recast in Methods. The first step needed is to spectrally flatten the Hamiltonian, which means that all positive eigenvalues of the Hamiltonian are replaced by +1, while all negative ones by -1. A crucial observation is that the spectral flattening actually amounts to replacing $\vec{h}(k)$ with a normalised one $\hat{h}(k)=\hat{h}_{x}(k)\hat{x}+\hat{h}_{y}(k)\hat{y}=\vec{h}(k)/|\vec{h}(k)|$. We can write
\begin{equation}
\frac{i}{4}c^{T}\bar{M}c=\sum_{k\in\text{BZ}}\phi_{k}^{\dagger}\left[\hat{h}(k)\cdot\vec{\sigma}\right]\phi_{k},
\end{equation}
where it is understood that the lefthand side is the conventional Majorana representation of the righthand side with $\bar{M}$ being a real skew-symmetric matrix. The next step is to project $c$ and $\bar{M}$ onto the subspace under consideration. For example, in order to obtain $\rho_{\text{a}_{n}\text{b}_{m}}$, one takes
\begin{equation}
\frac{i}{4}c_{\text{a}_{n}\text{b}_{m}}^{T}\bar{M}_{\text{a}_{n}\text{b}_{m}}c_{\text{a}_{n}\text{b}_{m}}=\frac{1}{N}\sum_{k\in\text{BZ}}\left(\begin{matrix}a_{n}^{\dagger}e^{ikn}\\b_{m}^{\dagger}e^{ikm}\end{matrix}\right)^{T}\hat{h}(k)\cdot\vec{\sigma}\left(\begin{matrix}a_{n}e^{-ikn}\\b_{m}e^{-ikm}\end{matrix}\right)=\eta_{\text{a}_{n}\text{b}_{m}}(a_{n}^{\dagger}b_{m}+b_{m}^{\dagger}a_{n}),
\label{eq:mab}
\end{equation}
where
\begin{align}
\eta_{\text{a}_{n}\text{b}_{m}}&=\frac{1}{N}\sum_{k\in\text{BZ}}\cos\left[k(m-n)+\varphi(k)\right],\\
e^{i\varphi(k)}&=\hat{h}_{x}(k)+i\hat{h}_{y}(k).
\end{align}
The quantity $\eta_{\text{a}_{n}\text{b}_{m}}$ will play a central role in our analysis. Once the $4\times4$ matrix $\bar{M}_{\text{a}_{n}\text{b}_{m}}$ is obtained from Eq.~\eqref{eq:mab}, one can follow the prescription in Methods to have
\begin{equation}
\rho_{\text{a}_{n}\text{b}_{m}}=\frac{1}{4}\left(\begin{matrix}
1-\eta^{2} & 0 & 0 & 0 \\
0 & 1+\eta^{2} & -2\eta & 0 \\
0 & -2\eta & 1+\eta^{2} & 0 \\
0 & 0 & 0 & 1-\eta^{2}
\end{matrix}\right)
\label{eq:reduced}
\end{equation}
in the basis $\{\ket{0},b_{m}^{\dagger}\ket{0},a_{n}^{\dagger}\ket{0},a_{n}^{\dagger}b_{m}^{\dagger}\ket{0}\}$, the subscript in $\eta_{\text{a}_{n}\text{b}_{m}}$ being omitted for brevity. 

\begin{figure}
\center\includegraphics[width=0.4\columnwidth]{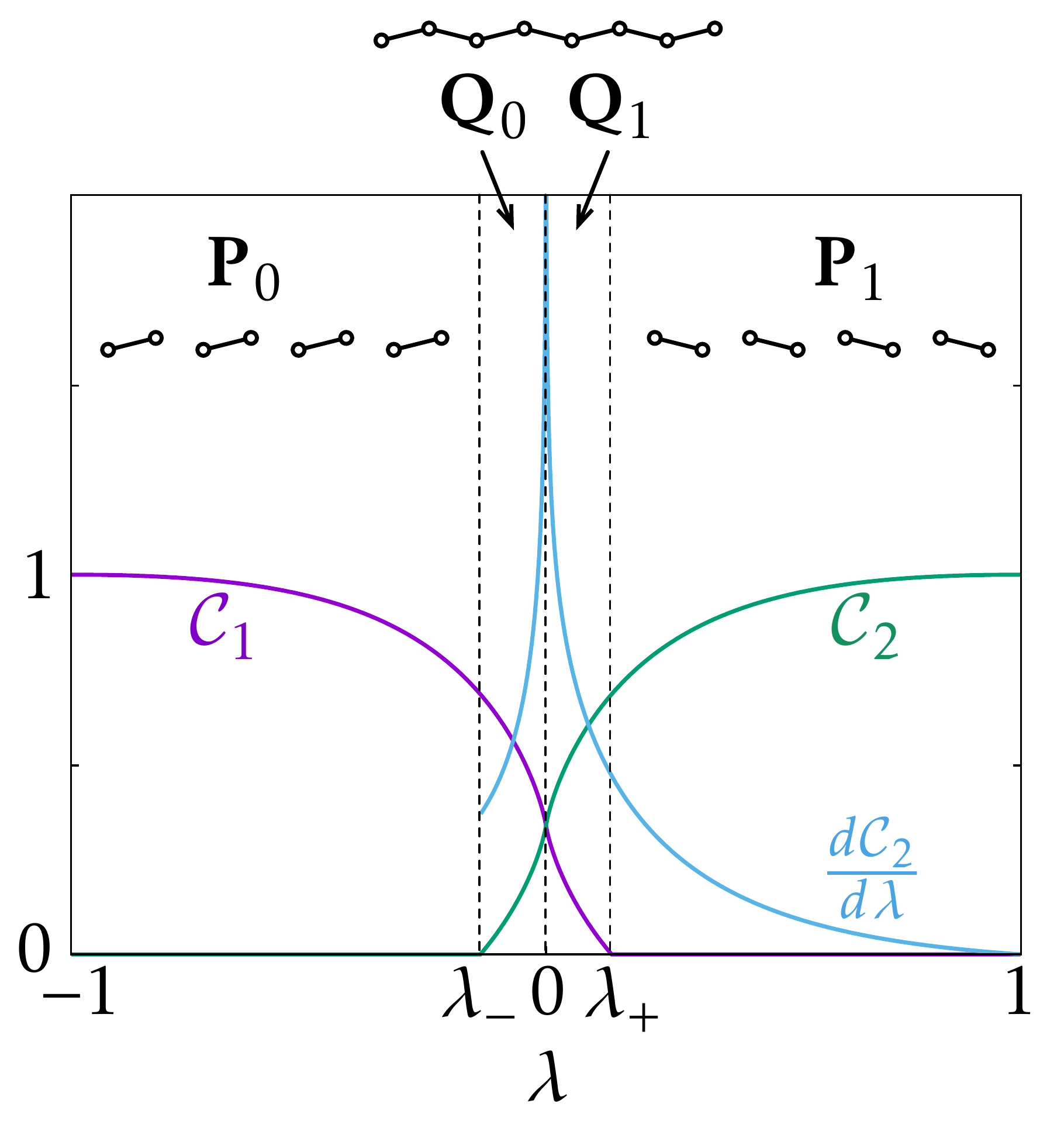}
\caption{Phase diagram of many-body entanglement in the thermodynamic limit. There exist four phases $\mathbf{P}_{\{0,1\}}$ and $\mathbf{Q}_{\{0,1\}}$ with transition points $\lambda_{0}=0$ and $\lambda_{\pm}\simeq\pm0.138$. The corresponding entangled graphs are depicted in the figure. $d\mathcal{C}_{1}/d\lambda$ is omitted as it can be obtained from $\mathcal{C}_{1}(\lambda)=\mathcal{C}_{2}(-\lambda)$.}
\label{fig:phase}
\end{figure}

\subsection*{Phase diagram of many-body entanglement}

The concurrence can be directly calculated from the reduced density matrix~\eqref{eq:reduced} as 
\begin{equation}
\mathcal{C}_{\text{a}_{n}\text{b}_{m}}=\max\left\{0, \frac{1}{2}(\eta_{\text{a}_{n}\text{b}_{m}}^{2}+2\eta_{\text{a}_{n}\text{b}_{m}}-1)\right\},
\label{eq:concurrence}
\end{equation}
which is nonzero only for $\eta_{\text{a}_{n}\text{b}_{m}}>\sqrt{2}-1$~\cite{hor09}. It turns out that $\eta_{\text{a}_{n}\text{a}_{m}}=\eta_{\text{b}_{n}\text{b}_{m}}=0$ for all $\{n,m\}$, which implies that $\rho_{\text{a}_{n}\text{a}_{m}}$ and $\rho_{\text{b}_{n}\text{b}_{m}}$ are fully mixed states and thus $\mathcal{C}_{\text{a}_{n}\text{a}_{m}}=\mathcal{C}_{\text{b}_{n}\text{b}_{m}}=0$. Furthermore, $\mathcal{C}(\text{a}_{n},\text{b}_{m})$ also all vanish except for $\mathcal{C}(\text{a}_{n},\text{b}_{n})$ and $\mathcal{C}(\text{b}_{n},\text{a}_{n+1})$. As a result, there are only three possible entangled graphs associated with the ground state, as shown in Figure~\ref{fig:phase}. Hereafter, for a notational simplicity, we let $\eta_{1}\equiv\eta_{\text{a}_{n}\text{b}_{n}}$, $\mathcal{C}_{1}\equiv\mathcal{C}_{\text{a}_{n}\text{b}_{n}}$, $\eta_{2}\equiv\eta_{\text{b}_{n}\text{a}_{n+1}}$, and $\mathcal{C}_{2}\equiv\mathcal{C}_{\text{b}_{n}\text{a}_{n+1}}$. The two concurrences are related as
\begin{equation}
\mathcal{C}_{1}(\lambda)=\mathcal{C}_{2}(-\lambda).
\label{eq:sym}
\end{equation}

In Figure~\ref{fig:phase}, we plot $\mathcal{C}_{1}(\lambda)$, $\mathcal{C}_{2}(\lambda)$, and $d\mathcal{C}_{2}(\lambda)/d\lambda$ with respect to $\lambda$. One can distinguish four different phases $\mathbf{P}_{\{0,1\}}$ and $\mathbf{Q}_{\{0,1\}}$. The subscript represents the $Z_{2}$ index for the band topology with 0 (1) representing the trivial (topological) phase. In the trivial (topological) phase, $\mathcal{C}_{1}>\mathcal{C}_{2}$ ($\mathcal{C}_{1}<\mathcal{C}_{2}$). In the phase $\mathbf{Q}$, the entangled graph is connected, while in the phase $\mathbf{P}$, it is disconnected. The phase transition of the band topology occurs at $\lambda_{0}=0$ and that of the entangled graph occurs at $\lambda_{\pm}\simeq\pm0.138$. 

\subsection*{Quantum phase transition at $\lambda=\lambda_{0}$}

\begin{figure}
\center\includegraphics[width=0.6\columnwidth]{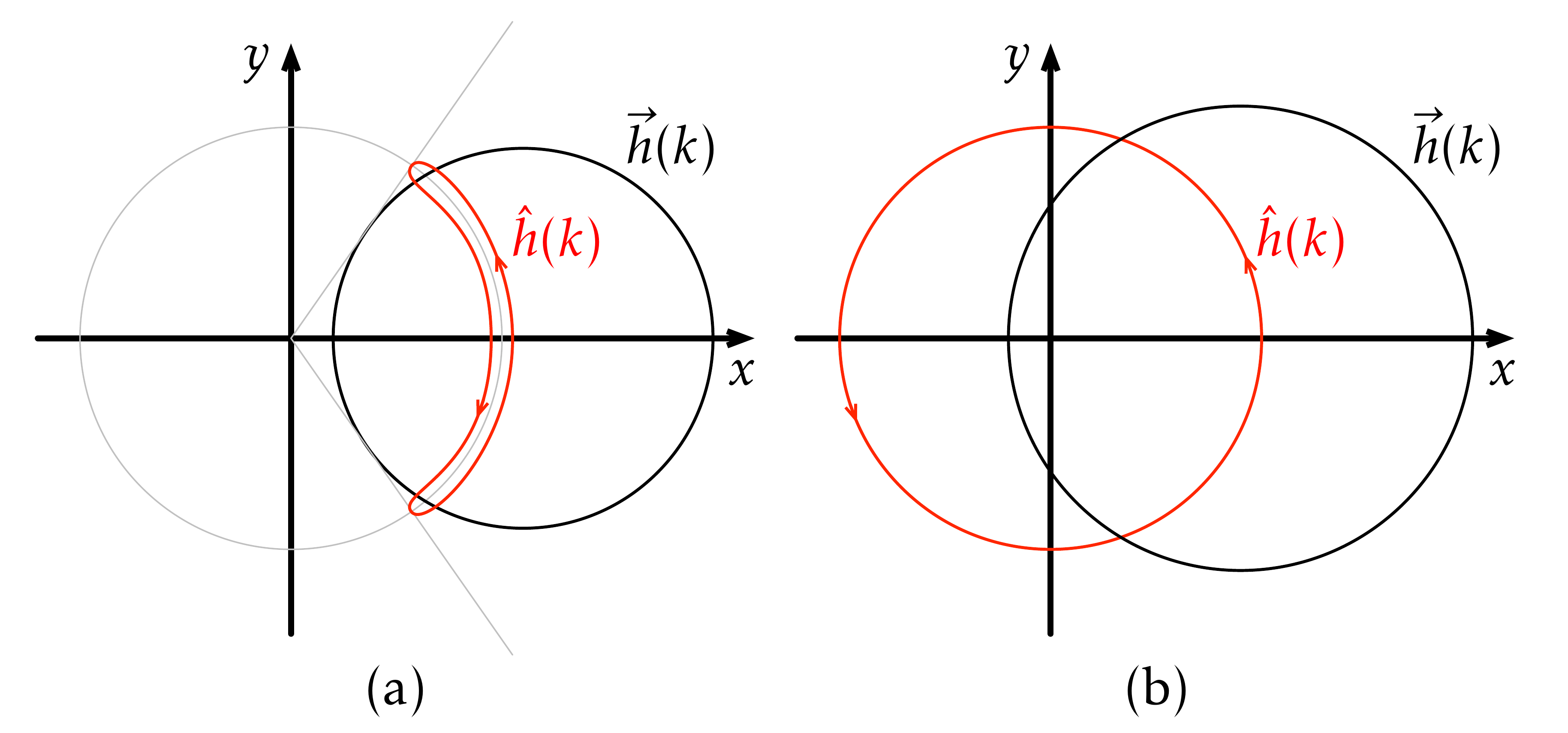}
\caption{The trajectories of $\vec{h}(k)$ and $\hat{h}(k)$ (a) in the topologically trivial phase ($\lambda<0$) and (b) in the topological phase ($\lambda>0$).}
\label{fig:topology}
\end{figure}

The nonanalyticity of the concurrence here can be ascribed to the sudden change of the band topology. The vector $\vec{h}(k)$ in Eq.~\eqref{eq:hamiltonian} traverses a circle on the $x-y$ plane as $k$ sweeps over the Brillouin zone $0\le k<2\pi$, as in Figure~\ref{fig:topology}. The $Z_{2}$ topological index is then determined by whether the circle encloses the origin or not~\cite{ali12}. In order to make this topological distinction more pronounced, one can take the unit vector $\hat{h}(k)$, as in Figure~\ref{fig:topology}. In the trivial phase ($\lambda<0$), $\hat{h}(k)$ wanders on the half circle in $x>0$. To close the loop, it traverses the same path twice. In the topological phase ($t_{1}<t_{2}$), on the other hand, $\hat{h}(k)$ traverses the full circle once. 

Such a topological difference is in fact captured by the quantity $\eta_{1}$:
\begin{equation}
\eta_{1}(\lambda)=\frac{1}{N}\sum_{k\in\text{BZ}}\hat{h}_{x}(k)\xrightarrow{N\rightarrow\infty}\frac{1}{2\pi}\int_{0}^{2\pi}\hat{h}_{x}(k)dk.
\label{eq:eta}
\end{equation}
At the phase transition, the trajectory of $\hat{h}(k)$ should change its shape from one to the other in Figure~\ref{fig:topology}. In the thermodynamic limit wherein the trajectory becomes {\em continuous}, such a change cannot be made continuously. In view of Eq.~\eqref{eq:eta}, this discontinuity should be reflected as a sudden jump of $\eta_{1}$ in the vicinity of the critical point, rendering the derivative of $\eta_{1}$ with respect to $\lambda$, and hence that of $\mathcal{C}_{1}$, diverging at the critical point. 

\begin{figure}
\center\includegraphics[width=0.96\columnwidth]{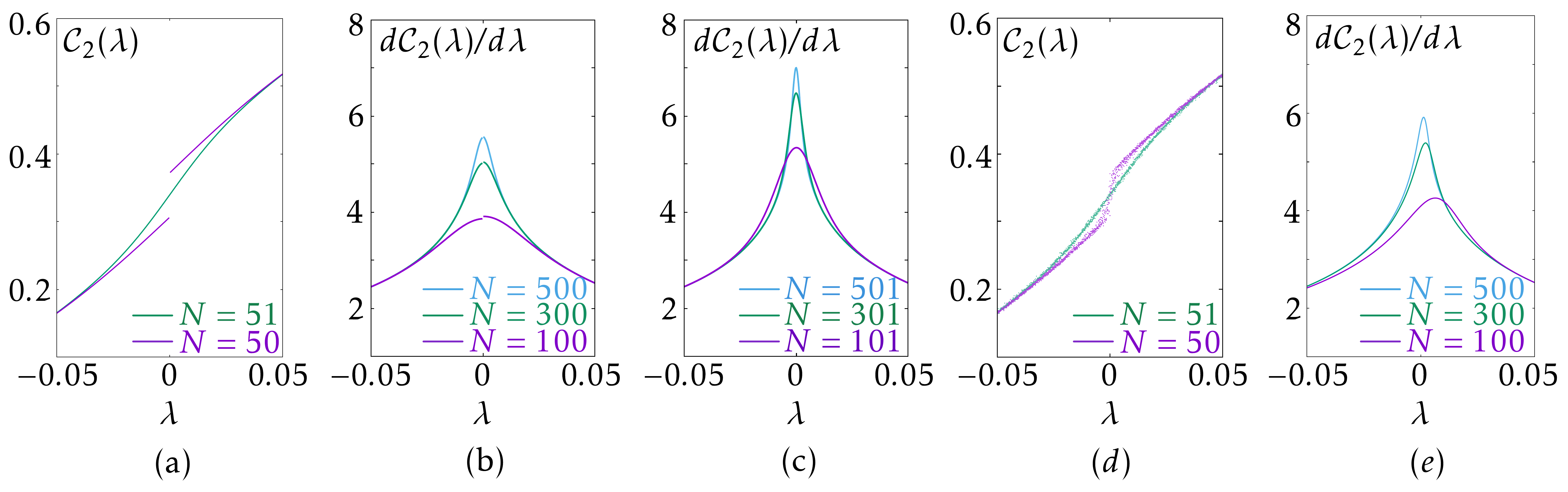}
\caption{(a) Different behaviour of $\mathcal{C}_{2}(\lambda)$ according to the parity of $N$. $\mathcal{C}_{2}(\lambda)$ is discontinuous at the critical point for even $N$. $d\mathcal{C}_{2}(\lambda)/d\lambda$ for (b) even $N$ and (c) odd $N$. (d) Typical behaviour of $\mathcal{C}_{2}(\lambda)$ in the presence of small disorder. (e) $d\mathcal{C}_{2}(\lambda)/d\lambda$ at the centre of the chain in the case of the open boundary condition.}
\label{fig:evenodd}
\end{figure}

This nonanalyticity, as shown in Figure~\ref{fig:phase}, is reminiscent of earlier results for the cases of symmetry-breaking quantum phase transitions in spin chains~\cite{ost02,osb02}. However, a stark difference is found for finite systems due to the different origins of the singularity. In the present case, the behaviour of $\eta_{1}(\lambda)$ depends on the parity of $N$, as shown in Figure~\ref{fig:evenodd}. For even $N$, $\eta_{1}(\lambda)$ is discontinuous at $\lambda=0$ even though the system size is finite. This can be understood by re-examining Figure~\ref{fig:topology} and Eq.~\eqref{eq:eta}. One can realise that for even $N$, as the trajectory of $\vec{h}(k)$ is made of $N$ equally spaced points on the circle, the difference between $\eta_{1}(0_{+})$ and $\eta_{1}(0_{-})$ is made solely by the contribution of $\hat{h}_{x}(k=\pi)$, which is either $1$ or $-1$ depending on the phase. We thus find that
\begin{equation}
\delta(N)\equiv\left|\eta_{1}(0_{+})-\eta_{1}(0_{-})\right|=\frac{2}{N}.
\end{equation}
For odd $N$, due to the exclusion of $k=\pi$, the discrete nature of $\eta_{1}(\lambda)$ disappears.

We perform additional calculations taking more realistic situations into account. \figurename~\ref{fig:evenodd}(d) shows $\mathcal{C}_{2}(\lambda)$ in the presence of disorder, which is introduced by adding $\sum_{n}(\epsilon_{n}^{a}a_{n}^{\dagger}a_{n}+\epsilon_{n}^{b}b_{n}^{\dagger}b_{n})$ to the Hamiltonian with $\epsilon_{n}^{\{a,b\}}$ being taken randomly and uniformly from the interval $[-0.1,0.1]$. It turns out that small disorder does not significantly alter the essential features of the entanglement including the distinction between the cases of even and odd $N$. \figurename~\ref{fig:evenodd}(e) shows $d\mathcal{C}_{2}/d\lambda$ obtained by taking the two sites at the centre of the chain in the case of the open boundary condition. In this case, the peak of $d\mathcal{C}_{2}/d\lambda$ does not coincide with $\lambda=0$ because the symmetry with respect to $\lambda\leftrightarrow-\lambda$, as in Eq.~\eqref{eq:sym}, is broken down. Also, the difference according to the parity of $N$ is absent because its origin---the inclusion or exclusion of $k=\pi$ in the Brillouin zone---now loses its meaning. Here, the chemical potential is placed in the gap either below or above the two zero-energy states that appear due to the open boundary condition. Either case produces the same results. As is expected, as $N$ increases, the singularity is more pronounced and the peak approaches $\lambda=0$.

It is worthwhile to interpret the nonanalyticity in terms of the conventional language of the phase transition. In fact, the derivative of the free energy with respect to $\lambda$ picks up $\eta$:
\begin{equation}
\begin{split}
\left.\frac{\partial}{\partial\lambda}F(T,\lambda)\right|_{T\rightarrow0}
&=N\avr{-a_{1}^{\dagger}b_{1}+b_{1}^{\dagger}a_{2}+\text{H.c.}}_{T\rightarrow0}\\
&=N\left[\eta_{1}(\lambda)-\eta_{2}(\lambda)\right],
\end{split}
\end{equation}
where $F(T,\lambda)=-k_{B}T\ln\text{Tr}[e^{-H(\lambda)/(k_{B}T)}]$. As the phase transition is second order, the derivative of $\eta$, and hence the derivative of  the concurrence, diverges at the critical point. In the thermodynamic limit, $\eta_{1}(\lambda)$ diverges logarithmically for $|\lambda|\ll1$ as 
\begin{equation}
\frac{\partial}{\partial\lambda}\eta_{1}(\lambda)\simeq\frac{2}{\pi}\log\left[\left(\frac{e}{2}\right)^{2}|\lambda|\right],
\label{eq:diverge}
\end{equation}
which can be derived by using the property of the elliptic integral: $-\int_{0}^{\pi/2}(\lambda^{2}+\sin^{2}\theta)^{-1/2}d\theta\simeq\log(\lambda/4)$. 

The above arguments remain valid for arbitrary one-dimensional topological phase transitions governed by a two-band Hamiltonian as in Eq.~\eqref{eq:hamiltonian}. In the thermodynamic limit, there should be a local quantity corresponding to $\eta_{1}$, which becomes nonanalytic at the critical point, reflecting the change of the winding number. For finite systems, it is discontinuous if the band-touching point belongs to the (discrete) Brillouin zone, which would be naturally related to a commensurability problem. For example, in the Kitaev chain~\cite{kit01}, the reduced density matrix for a single site, say, site 1, is obtained as
\begin{equation}
\rho_{1}=\frac{1}{N}\sum_{k\in\text{BZ}}\hat{h}_{z}(k)(a_{1}^{\dagger}a_{1}-a_{1}a_{1}^{\dagger}),
\end{equation}
where $a_{1}$ is the fermion operator for site 1. Note that the coefficient is identical to Eq.~\eqref{eq:eta} except for the orientation of the vector being different. In this case, the local electron density $\langle a_{1}^{\dagger}a_{1}\rangle$ as a function of the chemical potential is discontinuous at the critical point for finite even $N$ and the local compressibility diverges logarithmically as Eq.~\eqref{eq:diverge} in the thermodynamic limit. The latter behaviour has been addressed in a recent work~\cite{noz16}.

\subsection*{Phase transition of entanglement at $\lambda=\lambda_{\pm}$}

The nonanalyticity of the concurrence here is originated from the property of entanglement, which is defined as ``not being separable''. In the Hilbert space of bipartite density operators, the set of all separable states forms a compact convex set~\cite{hor96,ter00}. As a result, if one traces a continuous path in the Hilbert space from an entangled to a separable state, the entanglement suddenly disappears when one crosses the hyperplane separating the sets of separable and entangled states. When this occurs in a dynamical problem, the phenomenon is called an entanglement sudden death (or sudden birth in the opposite way)~\cite{yu09}. The entanglement sudden death is observed occasionally when a state evolves in a dissipative environment. However, it is rare to see an analogous phenomenon in the course of a quantum phase transition (to our best efforts, we could not find out a prior example). In view of Eq.~\eqref{eq:concurrence}, the existence of this phase transition should be robust against a small perturbation to the system. As the system remains gapped, one can transform it into a dynamical problem by considering an adiabatic evolution varying $\lambda$, making the link to the entanglement sudden death clearer. 

\begin{figure}
\center\includegraphics[width=0.4\columnwidth]{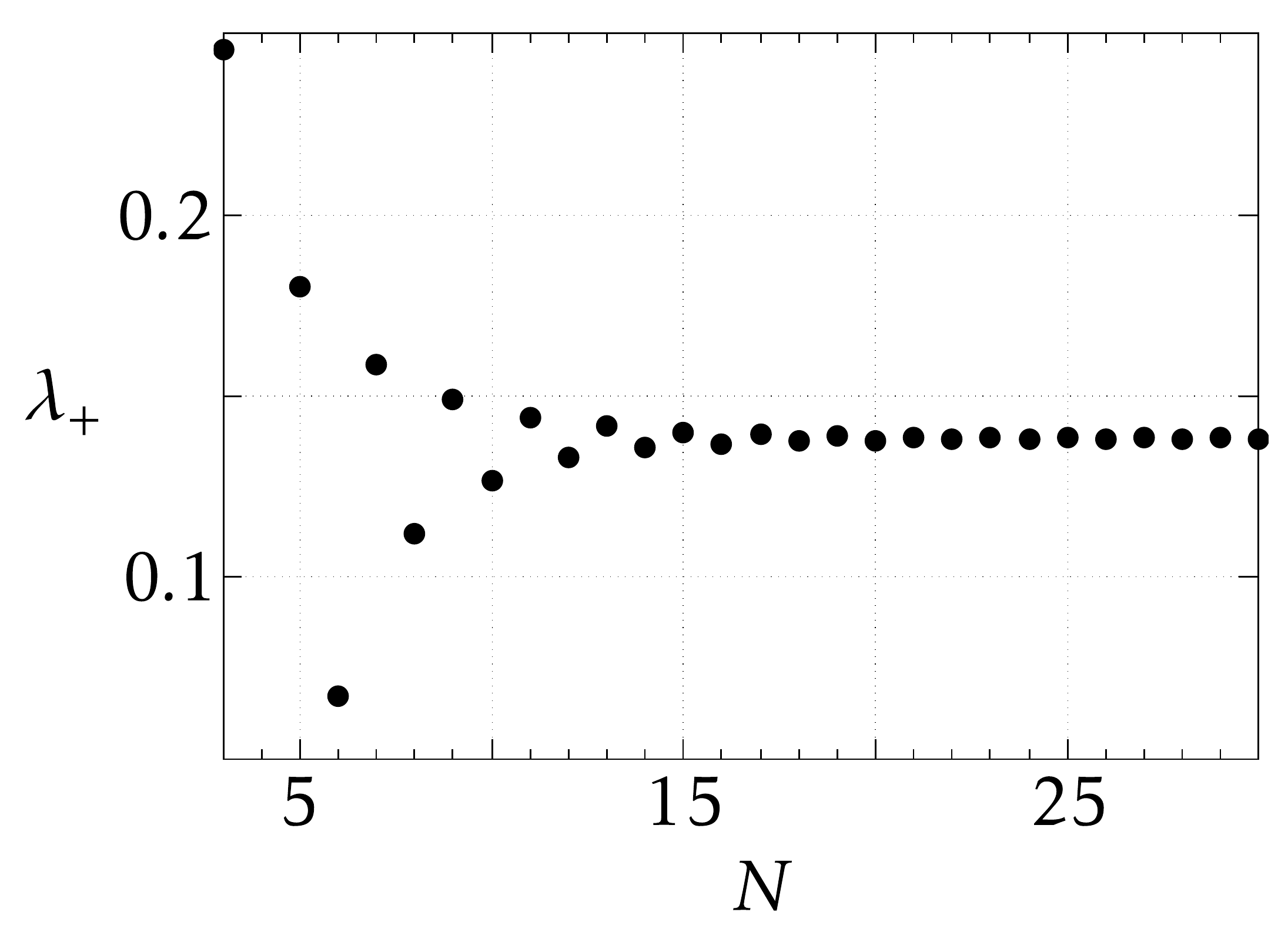}
\caption{Phase transition point $\lambda_{+}$ of the entangled graph for different system sizes.}
\label{fig:lambda}
\end{figure}

Near $\lambda=\lambda_{+}$, $\mathcal{C}_{1}(\lambda)$ changes linearly for $\lambda<\lambda_{+}$ and vanished for $\lambda>\lambda_{+}$. In the thermodynamic limit, the former behaviour is given by
\begin{equation}
\mathcal{C}_{1}(\lambda)\simeq-1.476(\lambda-\lambda_{+}).
\end{equation}
For finite $N$, $\lambda_{+}$ changes with $N$ as shown in \figurename~\ref{fig:lambda}. For $N=2$ or $N=4$, $\mathcal{C}_{1}>0$ for $\lambda<0$ and $\mathcal{C}_{1}=0$ for $\lambda>0$, and hence we cannot find $\lambda_{+}$.

As we have addressed in Introduction, this singularity is not originated from the nonanalyticity of the ground-state wave function and hence the transition is not a quantum phase transition in the conventional sense. Nonetheless, the transition may find meaning  in the context of quantum information theory. For example, supposing two parties, say, Alice and Bob, possess each of the two sites in a unit cell, the usefulness of the state as a resource for a quantum information processing depends on the phase~\cite{nie00}. Here, instead of considering entanglement of the fermion occupation number, which is somewhat impractical, one can make the scenario more practical by turning the SSH model into an equivalent $s=1/2$ spin-chain model with nearest-neighbour interaction. That is, regarding $a_{n}$ and $b_{n}$ in Eq.~\eqref{eq:model} as $s=1/2$ spin lowering operators, in the subspace with the half-filling $\sum_{n}(a_{n}^{\dagger}a_{n}+b_{n}^{\dagger}b_{n})=N$, one can realise that the equivalent spin-chain Hamiltonian is
\begin{equation}
H_{\text{spin}}=\sum_{n=1}^{N-1}(t_{1}a_{n}^{\dagger}b_{n}+t_{2}b_{n}^{\dagger}a_{n+1}+\text{H.c.})+t_{1}(a_{N}^{\dagger}b_{N}+\text{H.c.})+(-1)^{N-1}t_{2}(b_{N}^{\dagger}a_{1}+\text{H.c.}).
\end{equation}
Note that the last term introduces a gauge field for even $N$, which would be responsible for the different behaviour at $\lambda=\lambda_{0}$ for even and odd $N$. A deeper operational meaning of the entanglement phase transition at $\lambda=\lambda_{\pm}$ is open to question due to the lack of a relevant quantum informational protocol in the multipartite setting.

\section*{Discussion}

We have fully characterised two-site entanglements in the ground state of the SSH model and obtained the phase diagram of the entanglement. It was found out that there are two kinds of singularities in the entanglement: one due to the topological quantum phase transition and the other one due to the entanglement sudden death. 

Several remarks are in order. 
(i) The presented singularities are expected to be verifiable in optical lattices. For example, one could prepare a fermionic Mott insulator in a one-dimensional optical lattice and adiabatically change the potential shape by superimposing an additional lattice potential so that the number of sites is twice the number of atoms in the end~\cite{jor08,ata13}. More detailed and rigorous analysis of its feasibility is left as a future work.
(ii) As can be seen in Figure~\ref{fig:phase}, one can detect the topological order of the SSH model simply by comparing local quantities $\mathcal{C}_{1}$ and $\mathcal{C}_{2}$, instead of referring to the entanglement spectrum. This might greatly facilitate the experimental detection of the topological phase. However, the validity of this idea in the presence of disorder or interaction and its generalisation to other models are unclear. (iii) While our way of characterising the many-body entanglement seems reasonable for the SSH model, one can employ a different characterisation to see a different aspect of the many-body entanglement. The phases in Figure~\ref{fig:phase} would then be divided into subphases, enriching the phase diagram.

\section*{Methods}

\subsection*{Reduced density matrix for noninteracting gapped fermion systems}

We recast the formulae presented in Refs.~\cite{kit01,bra04,fid10} to derive a reduced density matrix of the ground state of a gapped quadratic fermion Hamiltonian. Consider a system described by $N$ fermion annihilation (creation) operators $a_{j}$ ($a_{j}^{\dagger}$) with $1\le j\le N$. It is convenient to introduce Majorana operators $c_{l}$ with $1\le l\le 2N$  such that $a_{j}=\frac12(c_{2j-1}+ic_{2j})$. Let $c$ be the vector with $2N$ elements $c_{l}$. Hereafter, we will use a similar convention for other operators as well. The Hamiltonian can be generally written as
\begin{equation}
H=\frac{i}{4}c^{T}Mc,
\end{equation}
where $M$ is a $2N\times2N$ real skew-symmetric matrix with $M_{jk}=-M_{kj}$. The matrix $M$ is block diagonalised as
\begin{equation}
WMW^{T}=\left(\begin{matrix}
0 & \epsilon_{1} & & & \\
-\epsilon_{1} & 0 & & & \\
& & \ddots & & \\
& & & 0 & \epsilon_{N} \\
& & & -\epsilon_{N} & 0
\end{matrix}\right)
\label{eq:blockd}
\end{equation}
by a $2N\times2N$ real orthogonal matrix $W$ with $W^{T}W=WW^{T}=I$, where $\epsilon_{j}>0$. The Hamiltonian is then written as
\begin{equation}
H=\frac{i}{2}\sum_{j=1}^{N}\epsilon_{j}c'_{2j-1}c'_{2j}=\sum_{j=1}^{N}\epsilon_{j}\left({a'_{j}}^{\dagger}a'_{j}-\frac{1}{2}\right),
\end{equation}
where $c'=Wc$ and $a'_{j}=\frac12(c'_{2j-1}+ic'_{2j})$. The ground state in the density matrix form is thus
\begin{equation}
\rho_{0}=\prod_{j=1}^{N}a'_{j}{a'_{j}}^{\dagger}=\prod_{j=1}^{N}\left(\frac{1}{2}-i\frac{1}{2}c'_{2j-1}c'_{2j}\right).
\end{equation}
For later use, let us define a spectral flattening of the Hamiltonian $H$ as
\begin{equation}
\bar{H}=\frac{i}{4}c^{T}\bar{M}c=\sum_{j=1}^{N}\left({a'_{j}}^{\dagger}a'_{j}-\frac{1}{2}\right),
\label{eq:flatten}
\end{equation}
which amounts to replacing all $\epsilon_{j}$ with 1. We also define a Grassmann representation $\omega(X)$ for a polynomial $X$ of Majorana operators, which is done by replacing all Majorana operators in $X$ with Grassmann variables. For example, a Grassmann representation of $\rho_{0}$ is obtained by replacing $c_{l}$ and $c'_{l}$ with Grassmann variables $\theta_{l}$ and $\theta'_{l}=\sum_{m}W_{lm}\theta_{m}$, respectively, as
\begin{equation}
\omega(\rho_{0})=\prod_{j=1}^{N}\left(\frac{1}{2}-i\frac{1}{2}\theta'_{2j-1}\theta'_{2j}\right)=\frac{1}{2^{N}}\exp\left(-\frac{i}{2}\theta^{T}\bar{M}\theta\right).
\label{eq:gaussian}
\end{equation}
Note that $\bar{M}$ is obtained from the spectrally flattened Hamiltonian~\eqref{eq:flatten}. 
A state that has a Gaussian form of a Grassmann representation as in Eq.~\eqref{eq:gaussian} is called a Gaussian state. It can be checked that $\bar{M}$ is in fact a two point correlation Matrix
\begin{equation}
\bar{M}_{jk}=\left\{\begin{array}{ll}
-i\text{Tr}(\rho_{0}c_{j}c_{k}) & \text{ for }j\not=k\\
0 & \text{ for }j=k\end{array}\right.
\end{equation}
and higher order correlations are given by
\begin{equation}
\text{Tr}(\rho_{0}c_{j_{1}}c_{j_{2}}\cdots c_{j_{2n}})=i^{n}\text{Pf}(\bar{M}|_{j_{1}j_{2}\cdots j_{2n}})
\end{equation}
for $j_{1}<j_{2}<\cdots<j_{2n}$, where $\bar{M}|_{j_{1}j_{2}\cdots j_{2n}}$ is a $2n\times2n$ submatrix of $\bar{M}$ restricted to the designated indices and $\text{Pf}(\cdot)$ is the Pfaffian. All odd-order correlations vanish. 

Now let us divide the system into two parts $A$ and $B$. $A$ is described by $N_{A}$ pairs of fermion operators and $B$ by the rest. Our aim is to calculate the reduced density matrix $\rho_{A}=\text{Tr}_{B}(\rho_{0})$. One can check that if the Grassmann representation of $\rho_{A}$ is given by
\begin{equation}
\omega(\rho_{A})=\frac{1}{2^{N_{A}}}\exp\left(-\frac{i}{2}\theta_{A}^{T}\bar{M}_{A}\theta_{A}\right),
\label{eq:grassmann}
\end{equation}
where $\theta_{A}$ and $\bar{M}_{A}$ are the corresponding submatrices restricted to the indices for the subsystem $A$, then for an arbitrary polynomial $X_{A}$ of the Majorana operators supported on A,
\begin{equation}
\text{Tr}(\rho_{A}X_{A})=\text{Tr}(\rho_{0}X_{A}).
\end{equation}
This implies that Eq.~\eqref{eq:grassmann} is indeed correct. As $\bar{M}_{A}$ is also skew-symmetric, it is block diagonalised as
\begin{equation}
W_{A}\bar{M}_{A}W_{A}^{T}=\left(\begin{matrix}
0 & \eta_{1} & & & \\
-\eta_{1} & 0 & & & \\
& & \ddots & & \\
& & & 0 & \eta_{N} \\
& & & -\eta_{N} & 0
\end{matrix}\right),
\end{equation}
where $W_{A}$ is real orthogonal and $0\le\eta_{j}\le1$. Taking the submatrix $c_{A}$ of $c$ and letting $c'_{A}=W_{A}c_{A}$, we finally obtain
\begin{equation}
\rho_{A}=\prod_{j=1}^{N_{A}}\left(\frac{1}{2}-i\frac{\eta_{j}}{2}c'_{A,2j-1}c'_{A,2j}\right),
\end{equation}
which has the Grassmann representation as in Eq.~\eqref{eq:grassmann}.

\section*{Acknowledgements}

We thank Gil Young Cho, Isaac Kim, Ki-Seok Kim, and Heung-Sun Sim for fruitful discussions. This research was supported (in part) by the R\&D Convergence Program of NST (National Research Council of Science and Technology) of Republic of Korea (Grant No. CAP-15-08-KRISS). K.W.K. acknowledges support from the ``Overseas Research Program for Young Scientists'' by Korea Institute for Advanced Study (KIAS).

\section*{Author contributions statement}

J.C. conceived the preliminary idea and then J.C. and K.W.K. equally contributed to the analysis. J.C. wrote the manuscript.

\section*{Additional Information}

\textbf{Competing financial interests:} The authors declare no competing financial interests.


\begin{thebibliography}{10}
\expandafter\ifx\csname url\endcsname\relax
  \def\url#1{\texttt{#1}}\fi
\expandafter\ifx\csname urlprefix\endcsname\relax\def\urlprefix{URL }\fi
\providecommand{\bibinfo}[2]{#2}
\providecommand{\eprint}[2][]{\url{#2}}

\bibitem{sac11}
\bibinfo{author}{Sachdev, S.}
\newblock \emph{\bibinfo{title}{Quantum Phase Transition}}
  (\bibinfo{publisher}{Cambridge University Press, Cambridge, England},
  \bibinfo{year}{2011}), \bibinfo{edition}{2} edn.

\bibitem{wen04}
\bibinfo{author}{Wen, X.-G.}
\newblock \emph{\bibinfo{title}{Quantum Field Theory of Many-Body Systems}}
  (\bibinfo{publisher}{Oxford University Press, Oxford, England},
  \bibinfo{year}{2004}).

\bibitem{wen90}
\bibinfo{author}{Wen, X.~G.} \& \bibinfo{author}{Niu, Q.}
\newblock \bibinfo{title}{Ground-state degeneracy of the fractional quantum
  hall states in the presence of a random potential and on high-genus riemann
  surfaces}.
\newblock \emph{\bibinfo{journal}{Phys. Rev. B}} \textbf{\bibinfo{volume}{41}},
  \bibinfo{pages}{9377--9396} (\bibinfo{year}{1990}).

\bibitem{kit01}
\bibinfo{author}{Kitaev, A.~Y.}
\newblock \bibinfo{title}{Unpaired majorana fermions in quantum wires}.
\newblock \emph{\bibinfo{journal}{Phys.-Usp.}} \textbf{\bibinfo{volume}{44}},
  \bibinfo{pages}{131} (\bibinfo{year}{2001}).

\bibitem{sch08}
\bibinfo{author}{Schnyder, A.~P.}, \bibinfo{author}{Ryu, S.},
  \bibinfo{author}{Furusaki, A.} \& \bibinfo{author}{Ludwig, A. W.~W.}
\newblock \bibinfo{title}{Classification of topological insulators and
  superconductors in three spatial dimensions}.
\newblock \emph{\bibinfo{journal}{Phys. Rev. B}} \textbf{\bibinfo{volume}{78}},
  \bibinfo{pages}{195125} (\bibinfo{year}{2008}).

\bibitem{gu09}
\bibinfo{author}{Gu, Z.-C.} \& \bibinfo{author}{Wen, X.-G.}
\newblock \bibinfo{title}{Tensor-entanglement-filtering renormalization
  approach and symmetry-protected topological order}.
\newblock \emph{\bibinfo{journal}{Phys. Rev. B}} \textbf{\bibinfo{volume}{80}},
  \bibinfo{pages}{155131} (\bibinfo{year}{2009}).

\bibitem{che10}
\bibinfo{author}{Chen, X.}, \bibinfo{author}{Gu, Z.-C.} \&
  \bibinfo{author}{Wen, X.-G.}
\newblock \bibinfo{title}{Local unitary transformation, long-range quantum
  entanglement, wave function renormalization, and topological order}.
\newblock \emph{\bibinfo{journal}{Phys. Rev. B}} \textbf{\bibinfo{volume}{82}},
  \bibinfo{pages}{155138} (\bibinfo{year}{2010}).

\bibitem{fid10}
\bibinfo{author}{Fidkowski, L.}
\newblock \bibinfo{title}{Entanglement spectrum of topological insulators and
  superconductors}.
\newblock \emph{\bibinfo{journal}{Phys. Rev. Lett.}}
  \textbf{\bibinfo{volume}{104}}, \bibinfo{pages}{130502}
  (\bibinfo{year}{2010}).

\bibitem{has10}
\bibinfo{author}{Hasan, M.~Z.} \& \bibinfo{author}{Kane, C.~L.}
\newblock \bibinfo{title}{Colloquium: Topological insulators}.
\newblock \emph{\bibinfo{journal}{Rev. Mod. Phys.}}
  \textbf{\bibinfo{volume}{82}}, \bibinfo{pages}{3045} (\bibinfo{year}{2010}).

\bibitem{fis98}
\bibinfo{author}{Fisher, M.~E.}
\newblock \bibinfo{title}{Renormalization group theory: Its basis and
  formulation in statistical physics}.
\newblock \emph{\bibinfo{journal}{Rev. Mod. Phys.}}
  \textbf{\bibinfo{volume}{70}}, \bibinfo{pages}{653--681}
  (\bibinfo{year}{1998}).

\bibitem{ost02}
\bibinfo{author}{Osterloh, A.}, \bibinfo{author}{Amico, L.},
  \bibinfo{author}{Falci, G.} \& \bibinfo{author}{Fazio, R.}
\newblock \bibinfo{title}{Scaling of entanglement close to a quantum phase
  transition}.
\newblock \emph{\bibinfo{journal}{Nature}} \textbf{\bibinfo{volume}{416}},
  \bibinfo{pages}{608--610} (\bibinfo{year}{2002}).

\bibitem{osb02}
\bibinfo{author}{Osborne, T.~J.} \& \bibinfo{author}{Nielsen, M.~A.}
\newblock \bibinfo{title}{Entanglement in a simple quantum phase transition}.
\newblock \emph{\bibinfo{journal}{Phys. Rev. A}} \textbf{\bibinfo{volume}{66}},
  \bibinfo{pages}{032110} (\bibinfo{year}{2002}).

\bibitem{vid03}
\bibinfo{author}{Vidal, G.}, \bibinfo{author}{Latorre, J.~I.},
  \bibinfo{author}{Rico, E.} \& \bibinfo{author}{Kitaev, A.}
\newblock \bibinfo{title}{Entanglement in quantum critical phenomena}.
\newblock \emph{\bibinfo{journal}{Phys. Rev. Lett.}}
  \textbf{\bibinfo{volume}{90}}, \bibinfo{pages}{227902}
  (\bibinfo{year}{2003}).

\bibitem{kit06}
\bibinfo{author}{Kitaev, A.} \& \bibinfo{author}{Preskill, J.}
\newblock \bibinfo{title}{Topological entanglement entropy}.
\newblock \emph{\bibinfo{journal}{Phys. Rev. Lett.}}
  \textbf{\bibinfo{volume}{96}}, \bibinfo{pages}{110404}
  (\bibinfo{year}{2006}).

\bibitem{lev06}
\bibinfo{author}{Levin, M.} \& \bibinfo{author}{Wen, X.-G.}
\newblock \bibinfo{title}{Detecting topological order in a ground state wave
  function}.
\newblock \emph{\bibinfo{journal}{Phys. Rev. Lett.}}
  \textbf{\bibinfo{volume}{96}}, \bibinfo{pages}{110405}
  (\bibinfo{year}{2006}).

\bibitem{li08}
\bibinfo{author}{Li, H.} \& \bibinfo{author}{Haldane, F. D.~M.}
\newblock \bibinfo{title}{Entanglement spectrum as a generalization of
  entanglement entropy: Identification of topological order in non-abelian
  fractional quantum hall effect states}.
\newblock \emph{\bibinfo{journal}{Phys. Rev. Lett.}}
  \textbf{\bibinfo{volume}{101}}, \bibinfo{pages}{010504}
  (\bibinfo{year}{2008}).

\bibitem{ali12}
\bibinfo{author}{Alicea, J.}
\newblock \bibinfo{title}{New directions in the pursuit of {Majorana} fermions
  in solid state systems}.
\newblock \emph{\bibinfo{journal}{Rep. Prog. Phys.}}
  \textbf{\bibinfo{volume}{75}}, \bibinfo{pages}{076501}
  (\bibinfo{year}{2012}).

\bibitem{hor09}
\bibinfo{author}{Horodecki, R.}, \bibinfo{author}{Horodecki, P.},
  \bibinfo{author}{Horodecki, M.} \& \bibinfo{author}{Horodecki, K.}
\newblock \bibinfo{title}{Quantum entanglement}.
\newblock \emph{\bibinfo{journal}{Rev. Mod. Phys.}}
  \textbf{\bibinfo{volume}{81}}, \bibinfo{pages}{865} (\bibinfo{year}{2009}).

\bibitem{su79}
\bibinfo{author}{Su, W.~P.}, \bibinfo{author}{Schrieffer, J.~R.} \&
  \bibinfo{author}{Heeger, A.~J.}
\newblock \bibinfo{title}{Solitons in polyacetylene}.
\newblock \emph{\bibinfo{journal}{Phys. Rev. Lett.}}
  \textbf{\bibinfo{volume}{42}}, \bibinfo{pages}{1698--1701}
  (\bibinfo{year}{1979}).

\bibitem{ple03}
\bibinfo{author}{Plesch, M.} \& \bibinfo{author}{Bu\u{z}ek, V.}
\newblock \bibinfo{title}{Entangled graphs: Bipartite entanglement in
  multiqubit systems}.
\newblock \emph{\bibinfo{journal}{Phys. Rev. A}} \textbf{\bibinfo{volume}{67}},
  \bibinfo{pages}{012322} (\bibinfo{year}{2003}).

\bibitem{dur00}
\bibinfo{author}{D\"{u}r, W.}, \bibinfo{author}{Vidal, G.} \&
  \bibinfo{author}{Cirac, J.~I.}
\newblock \bibinfo{title}{Three qubits can be entangled in two inequivalent
  ways}.
\newblock \emph{\bibinfo{journal}{Phys. Rev. A}} \textbf{\bibinfo{volume}{62}},
  \bibinfo{pages}{062314} (\bibinfo{year}{2000}).

\bibitem{bri01}
\bibinfo{author}{Briegel, H.~J.} \& \bibinfo{author}{Raussendorf, R.}
\newblock \bibinfo{title}{Persistent entanglement in arrays of interacting
  particles}.
\newblock \emph{\bibinfo{journal}{Phys. Rev. Lett.}}
  \textbf{\bibinfo{volume}{86}}, \bibinfo{pages}{910} (\bibinfo{year}{2001}).

\bibitem{yu09}
\bibinfo{author}{Yu, T.} \& \bibinfo{author}{Eberly, J.~H.}
\newblock \bibinfo{title}{Sudden {Death} of {Entanglement}}.
\newblock \emph{\bibinfo{journal}{Science}} \textbf{\bibinfo{volume}{323}},
  \bibinfo{pages}{598--601} (\bibinfo{year}{2009}).

\bibitem{noz16}
\bibinfo{author}{Nozadze, D.} \& \bibinfo{author}{Trivedi, N.}
\newblock \bibinfo{title}{Compressibility as a probe of quantum phase
  transitions in topological superconductors}.
\newblock \emph{\bibinfo{journal}{Phys. Rev. B}} \textbf{\bibinfo{volume}{93}},
  \bibinfo{pages}{064512} (\bibinfo{year}{2016}).

\bibitem{hor96}
\bibinfo{author}{Horodecki, M.}, \bibinfo{author}{Horodecki, P.} \&
  \bibinfo{author}{Horodecki, R.}
\newblock \bibinfo{title}{Separability of mixed states: necessary and
  sufficient conditions}.
\newblock \emph{\bibinfo{journal}{Phys. Lett. A}}
  \textbf{\bibinfo{volume}{223}}, \bibinfo{pages}{1--8} (\bibinfo{year}{1996}).

\bibitem{ter00}
\bibinfo{author}{Terhal, B.~M.}
\newblock \bibinfo{title}{Bell inequalities and the separability criterion}.
\newblock \emph{\bibinfo{journal}{Phys. Lett. A}}
  \textbf{\bibinfo{volume}{271}}, \bibinfo{pages}{319--326}
  (\bibinfo{year}{2000}).

\bibitem{nie00}
\bibinfo{author}{Nielsen, M.} \& \bibinfo{author}{Chuang, I.}
\newblock \emph{\bibinfo{title}{Quantum Computation and Quantum Information}}
  (\bibinfo{publisher}{Cambridge University Press, Cambridge, England},
  \bibinfo{year}{2000}).

\bibitem{jor08}
\bibinfo{author}{J\"ordens, R.}, \bibinfo{author}{Strohmaier, N.},
  \bibinfo{author}{G\"unter, K.}, \bibinfo{author}{Moritz, H.} \&
  \bibinfo{author}{Esslinger, T.}
\newblock \bibinfo{title}{A {Mott} insulator of fermionic atoms in an optical
  lattice}.
\newblock \emph{\bibinfo{journal}{Nature}} \textbf{\bibinfo{volume}{455}},
  \bibinfo{pages}{204--207} (\bibinfo{year}{2008}).

\bibitem{ata13}
\bibinfo{author}{Atala, M.} \emph{et~al.}
\newblock \bibinfo{title}{Direct measurement of the {Zak} phase in topological
  bloch bands}.
\newblock \emph{\bibinfo{journal}{Nature Phys.}} \textbf{\bibinfo{volume}{9}},
  \bibinfo{pages}{795--800} (\bibinfo{year}{2013}).

\bibitem{bra04}
\bibinfo{author}{Bravyi, S.}
\newblock \bibinfo{title}{Lagrangian representation for fermionic linear
  optics}.
\newblock \eprint{arXiv:quant-ph/0404180}.

\end{thebibliography}
\end{document}